\documentstyle[aps,prc,eqsecnum,epsfig,floats,preprint]{revtex}
\tightenlines

\begin{document}
\draft

\title{ Dileptons from a Quark Gluon Plasma with Finite Baryon Density}

\author{A. Majumder and C. Gale}

\address{
Physics Department, McGill University, 3600 University St.,\\ Montr{\'e}al, QC.
Canada H3A 2T8}

\date{ \today}

\maketitle

\begin{abstract}
We investigate the effects of a baryon-antibaryon asymmetry on 
the spectrum of
dileptons radiating from a quark gluon plasma. We demonstrate the existence of 
a new set of processes in this regime. The dilepton production rate from
the corresponding diagrams is
shown to be as important as that obtained from the usual quark-antiquark
annihilation.   
\end{abstract}


\section{INTRODUCTION}

Experiments are now underway at the Relativistic Heavy Ion Collider (RHIC) at 
Brookhaven to study nuclear collisions at very high energies. 
The hope is to produce a
plasma of deconfined quarks and gluons. This
 plasma is,
however, rather ephemeral and soon hadronizes into a cornucopia of mesons 
and baryons.
One thus needs an indirect means of deducing as to whether or not the plasma was
 produced
 in the history of a given collision. Various experimental signatures have
been proposed to this effect: J$/\Psi$ suppression \cite{mat86}, 
strangeness enhancement \cite{raf82}, 
dilepton spectra \cite {shu80,kaj86,dum93} etc. 
In this paper we calculate a new contribution to the spectrum of
dileptons ( i.e., $e^{+}e^{-}, \mu^{+}\mu^{-} $) emanating from a quark 
gluon plasma.

Early calculations of the dilepton radiation in the deconfined sector were concerned
with the process 
$q\bar{q} \rightarrow e^{+}e^{-}$ \cite {shu80,kaj86,dum93}. 
A recent calculation has estimated the effects 
of chemical non-equilibrium and of a
large gluon excess
 on dilepton spectra \cite{lin00}. There, a fugacity 
was introduced to account for chemical
 non-equilibrium. The function of this fugacity is essentially
  to change the gluon and quark numbers from their equilibrium values. 
Possible sources of dileptons, 
such as $q+\bar{q}$ annihilation, $q+g$ Compton scattering,
   and $g+g$ fusion had been investigated and at chemical and
   thermal equilibrium the spectrum was found to be dominated by 
   $q\bar{q} \rightarrow e^{+}e^{-}$, followed by 
   $qg \rightarrow qge^{+}e^{-}$ which is an order of magnitude lower,
   followed by $gg \rightarrow q\bar{q}e^{+}e^{-}$ which is lower than the first
   process by 3 orders of magnitude \cite{lin00}.

The aim of our work is to propose that, when there 
is an asymmetry in the populations of quarks
and antiquarks (i.e., a finite baryon chemical
potential) a new
set of diagrams actually arise. 
Using these we calculate a new contribution to the
3-loop photon self-energy. 
The various cuts of this self-energy contain higher loop contributions to
the usual processes of $q\bar{q}\rightarrow e^{+}e^{-}$, 
$qg\rightarrow qe^{+}e^{-}$, $q q\rightarrow q q e^{+}e^{-}$, and an
entirely new process: $gg \rightarrow e^{+}e^{-}$. We calculate the
contribution of this new channel to the differential production rate 
of back-to-back dileptons. It is finally shown that within reasonable 
values of parameters
this process may become larger than the differential rate from the
standard tree level $q\bar{q}\rightarrow e^{+}e^{-}$.     

Imagine a scenario where the plasma is not just heated vacuum, but actually 
displays an asymmetry between quarks and antiquarks. This asymmetry would eventually 
manifest itself as an asymmetry between the baryon antibaryon populations in the final
state. 
In plasma calculations 
this asymmetry may be achieved by the introduction of a quark chemical
potential $\mu_{q}$. For the sake of simplicity we will assume here 
that $\mu$ is the same for $u$, $d$, and $s$ quarks 
(we are assuming the plasma to contain three massless flavours).
 We
also assume that the chemical potential for gluons is zero. The rest of the paper is
organised as follows: section II discusses a class of diagrams which are
non-existent at zero temperature, and also at
finite temperature and zero density. These become
finite at finite density. Section III focuses on a specific channel which will become a
source of dileptons. In section IV we derive a new contribution to the photon self-energy at
three loops and discuss its various cuts. Finally, in section V we calculate 
the production rate of back
to back dileptons.
All our calculations are done in the imaginary time formalism of equilibrium 
thermal field 
theory (our notation is described in Appendix A). 
We have assumed the presence of only three massless flavours of quarks. 
       
\vspace{1cm}
 
\section{NEW DIAGRAMS FROM BROKEN CHARGE CONJUGATION INVARIANCE }

\begin{figure}[htbp]
  \begin{center}
  \epsfxsize 120mm
  \epsfbox{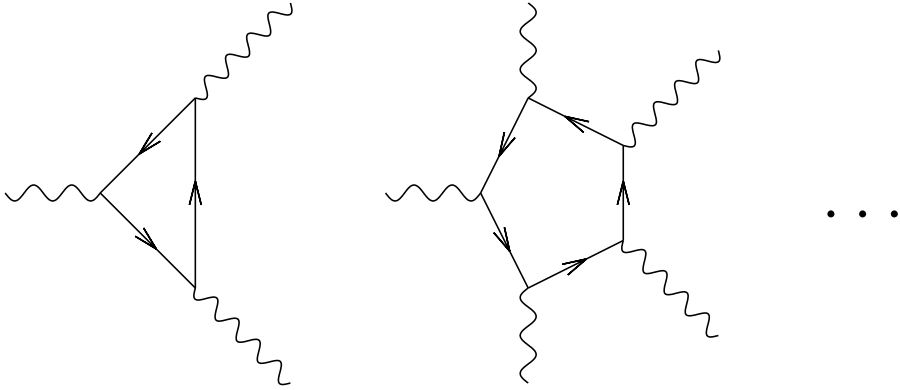}
\vspace{0.25cm}
    \caption{ Diagrams that are zero by Furry's theorem and extensions
thereof at finite temperature. These become non-zero at finite density.  }
    \label{furry}
  \end{center}
\end{figure}

At zero
temperature, and at finite temperature and zero baryon density,
diagrams in QED that contain a fermion loop with an odd number of
photon vertices (e.g. Fig. \ref{furry}) are cancelled by an equal and opposite 
contribution coming from the same diagram with fermion lines running in
the opposite direction (Furry's theorem \cite{itz80,wei95}). This statement can 
also be generalized to QCD for processes with two gluons and an odd 
number of photon vertices. 

However, at finite density or
 alternatively at finite fermion chemical potential, this cancellation no longer
occurs. As an illustration consider the diagrams of 
Fig. \ref{2vert}
for the case of two gluons and a photon attached to a quark loop 
(the analysis is the
same even for QED i.e., for three photons connected to an electron loop). 
In order
to obtain the full matrix element of a process containing the above as a 
sub-diagram one 
must coherently sum contributions from both diagrams which have fermion number
running in opposite directions. The amplitude for ${\mathcal T}^{\mu \rho \nu}   
( =  T^{\mu \rho \nu} +  T^{\nu \rho \mu} )$  are :

\begin{eqnarray}
\!\!\!\!\!T^{\mu \rho \nu} &=& \frac{1}{\beta} \sum_{n= -\infty}^{\infty}
 \int^{\infty}_{-\infty} eg^{2} tr[t^{a}t^{b}]
\frac{d^{3}q}{(2\pi)^{3}} Tr[ \gamma^{\mu} \gamma^{\beta} \gamma^{\rho} 
\gamma^{\delta} \gamma^{\nu} \gamma^{\alpha}]  \nonumber \\  
\mbox{} & & \frac{(q+p-k)_{\alpha} q_{\beta} 
(q+p)_{\delta} }{(q+p-k)^{2} q^{2} (q+p)^{2}}, \nonumber
\end{eqnarray}    

\noindent and

\begin{eqnarray}
\!\!\!\!\!T^{\nu \rho \mu} &=& \frac{1}{\beta} \sum_{n= -\infty}^{\infty}
 \int^{\infty}_{-\infty} eg^{2} tr[t^{a}t^{b}]
\frac{d^{3}q}{(2\pi)^{3}} Tr[ \gamma^{\nu} \gamma^{\delta} \gamma^{\rho} 
\gamma^{\beta} \gamma^{\mu} \gamma^{\alpha}] \nonumber \\ 
\mbox{} & & \frac{(q+k-p)_{\alpha} q_{\beta} 
(q-p)_{\delta} }{(q+k-p)^{2} q^{2} (q-p)^{2}}. \label{2g1p}
\end{eqnarray}

At finite temperature ($T$) and density (chemical potential $\mu$), 
we have the zeroth component of the fermion momenta given by, 

\begin{equation}
q_{0} = i (2n+1) \pi T + \mu  \hspace{1cm} \forall \; n \in I \label{q0}.
\end{equation}

We assume that $\mu$ is the same for both flavours of quarks.
Note that the extension of Furry's theorem to finite temperature does not hold at
finite density: as, if we set $n \rightarrow -2n-2$
we note that $ q_{0} \rightarrow \!\!\!\!\!\!/  -q_{0} $ and as a result

\begin{equation}
 T^{\mu \rho \nu} (\mu,T) \not= -T^{\nu \rho \mu} (\mu,T).
\end{equation}

\noindent Of course, If we now let the chemical potential go to zero 
($\mu \rightarrow 0$) we note
that for the transformation  $n \rightarrow -2n-2$  we obtain 
$q_{0} \rightarrow  -q_{0}$ and  thus 
$ T^{\mu \rho \nu} (0,T)  \rightarrow -T^{\nu \rho \mu} (0,T) $. The analysis for
fermion loops with larger number of vertices is essentially the same. 

 A more physical perspective is obtained by noting that all these 
diagrams are are encountered in the perturbative evaluation of Green's 
functions with an odd  number of gauge field operators. At zero 
(finite) temperature, in the well defined
 case of QED we observe quantities like 
$\langle 0| A_{\mu1} A_{\mu2} ... A_{\mu2n+1}
 |0\rangle $ ( $ Tr [ \rho(\mu,\beta)  A_{\mu1} A_{\mu2} ... A_{\mu2n+1} ] 
 $ ) under the action of the charge conjugation operator $C$. In QED we know that
 $CA_{\mu}C^{-1} = -A_{\mu} $. In the case of the
 vacuum $|0\rangle $, we note that $C|0\rangle = |0\rangle$, as the vacuum is 
uncharged. As a result

\begin{eqnarray}
 \langle 0| A_{\mu1} A_{\mu2} ... A_{\mu2n+1} |0\rangle &=&  
 \langle 0| C^{-1}C A_{\mu1} C^{-1}C A_{\mu2} ... A_{\mu2n+1} C^{-1} C |0\rangle
 \nonumber \\
  &=& \langle 0| A_{\mu1} A_{\mu2} ... A_{\mu2n+1} |0\rangle (-1)^{2n+1} 
\nonumber \\
&=&  -\langle 0| A_{\mu1} A_{\mu2} ... A_{\mu2n+1} |0\rangle. 
\end{eqnarray}

\noindent Hence, all such Green's functions are zero. The argument is the same for 
the case of
finite temperature and zero density (or chemical potential).  In the case of
nonzero chemical potential, however, the system is charged, hence the eigenstates of 
the
density operator $\rho$ are not eigenstates of $C$ i.e.,
 $C|n\rangle \not= |n\rangle$.
The medium, being charged, manifestly breaks charge conjugation invariance and these
 Green's functions are thus finite. The appearance of processes that can be related to 
 symmetry-breaking in a medium 
  has been noted before. \cite{chi77wel92}.


\section{THE TWO-GLUON-PHOTON VERTEX}

\begin{figure}[htbp]
  \begin{center}
  \epsfxsize 130mm
  \epsfbox{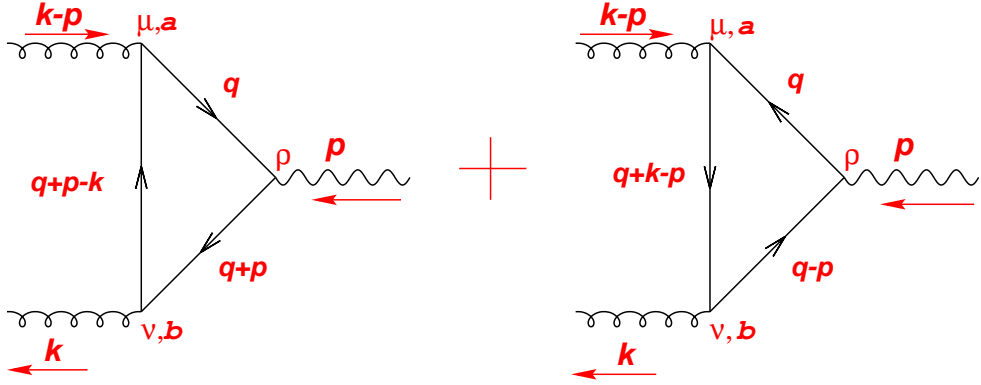}
\vspace*{0.3cm}
    \caption{ The two gluon photon effective vertex as the sum of two diagrams
with quark number running in opposite directions. }
    \label{2vert}
  \end{center}
\end{figure}

Let us now focus our attention on the diagrams of Fig. \ref{2vert}. Such a 
process does not exist at zero temperature or even at finite temperature and zero
 density. At finite density this may lead to a new source of dilepton or photon
 production. In this section we obtain general expressions for this diagram.

We sum the Matsubara frequencies using Pisarski's non-covariant method \cite{pis88} 
(Our notation is however different from
that in \cite{pis88} and is explained in Appendix A.). In this method,
propagators may be expanded as follows,

\begin{equation}
\frac{1}{q^{2}} =  \tilde{\Delta}(q)  = (-i) \int_{0}^{-i\beta} dx_{2}^{0} \; 
e^{-i(q^{0}-\mu)
x_{2}^{0} } \; \tilde{\Delta}^{\prime} (|\vec{q}| , -x_{2}^{0}), \label{t1}
\end{equation}

\noindent where

\[
\tilde{\Delta}^{\prime} (|\vec{q}| , -x_{2}^{0}) = \frac{1}{2E_{q}}
 \sum_{s_{2}} 
\tilde{f}_{s_{2}}^{\prime} (E_{q},s_{2}\mu) e^{-is_{2}(E_{q}+s_{2}\mu)x_{2}^{0}}, \nonumber 
\]

\noindent where $\tilde{f}^{\prime}_{+} = 1 - \tilde{n}(E+\mu)\; 
, \tilde{f}^{\prime}_{-} = - \tilde{n}(E-\mu)$, 
and $\tilde{n}$ is the Fermi-Dirac distribution function. Then, 

\begin{eqnarray}
\frac{1}{(q+p-k)^{2}} &=& \tilde{\Delta}(q+p-k) \nonumber \\
\mbox{} &=& (-i) \int_{0}^{-i\beta} dx_{1}^{0} \; 
 e^{i(q^{0}+p^{0}-k^{0}-\mu) x_{1}^{0} } \; \Delta (|\vec{q} + \vec{p} 
- \vec{k} |, x_{1}^{0}) \label{t2}
\end{eqnarray}

\[
\tilde{\Delta} (|\vec{q} + \vec{p} 
- \vec{k} | , x_{1}^{0}) = \frac{1}{2E_{q+p-k}}
 \sum_{s_{1}} 
\tilde{f}_{s_{1}} (E_{q+p-k},s_{1}\mu) e^{-is_{1}(E_{q+p-k}
-s_{1}\mu)x_{1}^{0}}, \nonumber
\]

\noindent where $\tilde{f}_{+} = 1 - \tilde{n}(E-\mu)\; 
, \tilde{f}_{-} = - \tilde{n}(E+\mu)$. For the last propagator, we may write

\begin{eqnarray}
\frac{1}{(q+p)^{2}} &=& \tilde{\Delta}(q+p) \nonumber \\
\mbox{} &=& (-i) \int_{0}^{-i\beta} dx_{3}^{0}\; 
e^{i(q^{0}+p^{0}-\mu) x_{3}^{0} } \; \tilde{\Delta} 
(|\vec{q} + \vec{p} |, x_{3}^{0}) \label{t3}
\end{eqnarray}

\noindent with 
\[
\tilde{\Delta} (|\vec{q} + \vec{p} | , x_{3}^{0}) = \frac{1}{2E_{q+p}}
 \sum_{s_{3}} 
\tilde{f}_{s_{3}} (E_{q+p},s_{3}\mu) e^{-is_{3}(E_{q+p}
-s_{3}\mu)x_{3}^{0}}. \nonumber
\]

\noindent In the above three equations $s_{1},s_{2},s_{3}$ are sign factors which 
are either +1 or -1,
and $q^{0} \rightarrow i( \frac{\partial}{\partial x_{2}^{0}} 
- i \mu )$. 
We substitute 
the above three equations in Eq. (\ref{2g1p}) and add the two diagrams. 
Now, one may simply 
perform the Matsubara sum to obtain a delta function over the times (see Appendix B). 
Using the 
delta function we may evaluate the remaining time integrations to give the 
full expression for
this vertex as (see Appendix B)

\begin{eqnarray}
\lefteqn{ {\mathcal T}^{\mu \rho \nu} = \int \frac{d^{3}q}{(2\pi)^{3}} 
\frac{\delta^{a b}}{2} e g^{2} \sum_{s_{1} s_{2} s_{3}}
Tr[\gamma^{\mu} \gamma^{\beta} \gamma^{\rho} 
\gamma^{\delta} \gamma^{\nu} \gamma^{\alpha}] 
\frac{\widehat{(q+p-k)}_{s_{1},\alpha} \hat{q}_{s_{2},\beta} 
\widehat{(q+p)}_{s_{3},\delta}}{p^{0}
+ s_{2}E_{q} - s_{3}E_{q+p}}} \nonumber \\ 
& & \!\!\!\!\!\!\!\!\!\!\!\!\!\!\!\!\!\!\!\!\!\!\left [ s_{2} \frac{ s_{3} 
(\tilde{n}(E_{q+p-k}+s_{1}\mu) - \tilde{n}(E_{q+p-k}-s_{1}\mu) ) -
s_{1}( \tilde{n}(E_{q+p}+s_{3}\mu) - \tilde{n}(E_{q+p}-s_{3}\mu) )}{k^{0} +
 s_{1}E_{q+p-k}
- s_{3}E_{q+p}} \right. \nonumber \\
& & \!\!\!\!\!\!\!\!\!\!\!\!\!\!\!\!\!\!\!\!\!\!\!\!\!
\left. \mbox{} - s_{3} \frac{s_{2}(\tilde{n}(E_{q+p-k}+s_{1}\mu) - 
\tilde{n}(E_{q+p-k}-s_{1}\mu)) -
s_{1}(\tilde{n}(E_{q}+s_{2}\mu) - \tilde{n}(E_{q}-s_{2}\mu)) }{k^{0}-p^{0} + 
s_{1}E_{q+p-k} - s_{2}E_{q}} \right]. \label{2g1pfv}
\end{eqnarray}

\noindent
In the above equation $E_{q-k} = \sqrt{ |\vec{q}|^{2} + |\vec{k}|^{2} - 
2|\vec{q}||\vec{k}|\cos{\theta} }$, where $\cos{\theta}$ is the angle between
$\vec{q}$ and $\vec{k}$. While in the numerator 

\[
\hat{q}_{s_{2}} = ( s_{2}, \sin{\theta}\cos{\phi} , \sin{\theta}\sin{\phi} , 
\cos{\theta} ),
\]

\noindent
and
  
\[
\widehat{(q-k)}_{s_{1}} = \Bigg( s_{1}, \frac{q\sin{\theta}\cos{\phi}}
{| \vec{q} - \vec{k} |} ,
\frac{q\sin{\theta}\sin{\phi}}{| \vec{q} - \vec{k} |} , 
\frac{q\cos{\theta} - k}{| \vec{q} - \vec{k} |} \Bigg).
\]

Up to this point no approximations have been made. Completing the remaining angular
integrations will lead to the expression for the effective vertex with arbitrary
momenta $k,p$. This, however, 
turns 
out to be rather difficult. Even in the Hard Thermal Loop (HTL) approximation,
 evaluation of similar three point loops for arbitrary three momentum $\vec{p}$ 
 is a difficult problem \cite{won92}. For the
HTL approximation to be valid, the temperature should be high, 
and the momentum
in the outer legs should be low. Predicted, initial temperatures at RHIC and 
LHC range from
300 to 800 MeV \cite{wan96,rap00}.  Dileptons from the
plasma are predicted to become important in the intermediate mass regime 
(1-3 GeV)\cite{shu80}. 
In this region of parameter space the HTL approximation obviously does not apply
. Hence we
do not make this approximation in this work. Furthermore, in the interest of 
technical simplicity, 
we proceed to evaluate the above diagram in the 
limit of the photon three momentum $\vec{p}=0$, i.e., for dilepton pairs produced
back-to-back. 

In this limit the expression for the effective vertex reduces to  

\begin{eqnarray}
{\mathcal T}^{\mu \rho \nu} &=& \int \frac{d^{3}q}{(2\pi)^{3}} 
\frac{\delta^{a b}}{2} e g^{2} \sum_{s_{1} s_{2} s_{3}}
Tr[\gamma^{\mu} \gamma^{\beta} \gamma^{\rho} 
\gamma^{\delta} \gamma^{\nu} \gamma^{\alpha}] 
\frac{\widehat{(q-k)}_{s_{1},\alpha} \hat{q}_{s_{2},\beta} 
\hat{q}_{s_{3},\delta}}{p^{0}
+ s_{2}E_{q} - s_{3}E_{q}} \nonumber \\ 
& & \!\!\!\!\!\!\!\!\!\!\!\!\!\!\!\!\!\!\!\!\!\!\left [ s_{2} \frac{ s_{3} 
 \Delta \tilde{n}(E_{q-k},s_{1}\mu)  -
s_{1} \Delta \tilde{n}(E_{q},s_{3}\mu) }{k^{0} +
 s_{1}E_{q-k}
- s_{3}E_{q}} 
 - s_{3} \frac{s_{2}\Delta\tilde{n}( E_{q-k},s_{1}\mu )  -
s_{1} \Delta\tilde{n}( E_{q},s_{2}\mu )  }{k^{0}-p^{0} + 
s_{1}E_{q-k} - s_{2}E_{q}} \right].  \label{2g1pb2b}
\end{eqnarray}

\noindent
 In the above, the
notation $ \Delta \tilde{n}(E,s\mu) = \tilde{n}(E+s\mu) - \tilde{n}(E-s\mu) $) 
is introduced.
The $ \Delta \tilde{n}$'s quantify the effect of a finite chemical potential, as
they represent the difference between the distribution functions of a quark and its
antiquark. If $\mu \rightarrow 0$, all the $ \Delta \tilde{n}$'s go to zero and so
does the  entire expression. 
Noting that $ s_{2} \Delta \tilde {n} (E_{q},s_{3}\mu) = s_{3} \Delta \tilde {n}
(E_{q},s_{2}\mu)$, we may add up the two terms in the integrand to give,

\begin{eqnarray}
{\mathcal T}^{\mu \rho \nu} &=& \int \frac{d^{3}q}{(2\pi)^{3}} 
\frac{\delta^{a b}}{2} e g^{2} \sum_{ s_{1} s_{2} s_{3} }
Tr[\gamma^{\mu} \gamma^{\beta} \gamma^{\rho} 
\gamma^{\delta} \gamma^{\nu} \gamma^{\alpha}] 
\widehat{(q-k)}_{s_{1},\alpha} \hat{q}_{s_{2},\beta} 
\hat{q}_{s_{3},\delta} \nonumber \\ 
& & \left[ 
\frac{ s_{2} s_{1} \Delta \tilde{n}( E_{q},s_{3}\mu ) - 
s_{2} s_{3} \Delta \tilde{n}( E_{q-k},s_{1}\mu ) }{ 
( k^{0} + s_{1}E_{q-k} - s_{3}E_{q} )
( k^{0}-p^{0} + s_{1}E_{q-k} - s_{2}E_{q} ) } 
  \right].   \label{2g1pcom}
\end{eqnarray}    
 
\noindent
Note that the integrand of the above expression for the vertex 
has factorized into a function of the form $f_{1}(k_{0})f_{2}(p_{0}-k_{0})$. 
This result will become important in the eventual evaluation of the 
photon self-energy in the subsequent sections.

We note that the sole $\phi$ dependence is in the numerator. This allows us to
integrate out the $\phi$ dependence. The presence of the Lorentz indices 
$ \alpha, \beta, \gamma $ indicate that Eq. (\ref{2g1pcom}) actually 
contains 64
separate terms. The $Tr[\gamma^{\mu} \gamma^{\beta} \gamma^{\rho} 
\gamma^{\delta} \gamma^{\nu} \gamma^{\alpha}]$ combines these into 64 other
terms. Ignoring the trace part, the rest of the formula is found to contain 
only 20 nonzero terms (see Appendix C).  
Of these 20 terms we observe many to be interdependent. There are at most 8
independent terms (see Appendix C).   

To perform the $\theta$ integration we shift $\vec{q}$ in the second term of the
integrand of Eq. (\ref{2g1pcom}) to give 

\begin{eqnarray}
{\mathcal T}^{\mu \rho \nu} &=& \int \frac{dq q^2 d\theta d\phi \sin{\theta} }
{(2\pi)^{3}} 
\frac{\delta^{a b}}{2} e g^{2} \sum_{ s_{1} s_{2} s_{3} }
Tr[\gamma^{\mu} \gamma^{\beta} \gamma^{\rho} 
\gamma^{\delta} \gamma^{\nu} \gamma^{\alpha}] 
 \nonumber \\ 
& & \left[ \widehat{(q-k)}_{s_{1},\alpha} \hat{q}_{s_{2},\beta} 
\hat{q}_{s_{3},\delta}
\frac{ s_{2} s_{1} \Delta \tilde{n}( E_{q},s_{3}\mu )  }{ 
( k^{0} + s_{1}E_{q-k} - s_{3}E_{q} )
( k^{0}-p^{0} + s_{1}E_{q-k} - s_{2}E_{q} ) } \right. \nonumber \\
& & \mbox{} - \left.
\hat{q}_{s_{1},\alpha} \widehat{(q+k)}_{s_{2},\beta} 
\widehat{(q+k)}_{s_{3},\delta}
\frac{ s_{2} s_{3} \Delta \tilde{n}( E_{q},s_{1}\mu ) }{
( k^{0} + s_{1}E_{q} - s_{3}E_{q+k} )
( k^{0}-p^{0} + s_{1}E_{q} - s_{2}E_{q+k} ) }
  \right].   \label{2g1pshft}
\end{eqnarray}    

\noindent 
The resulting $(\theta,\phi)$ integration now becomes rather simple.

\vspace{1cm}

\section{THE PHOTON SELF-ENERGY AND ITS IMAGINARY PART}

\begin{figure}[htbp]
  \begin{center}
  \epsfxsize 80mm
  \epsfbox{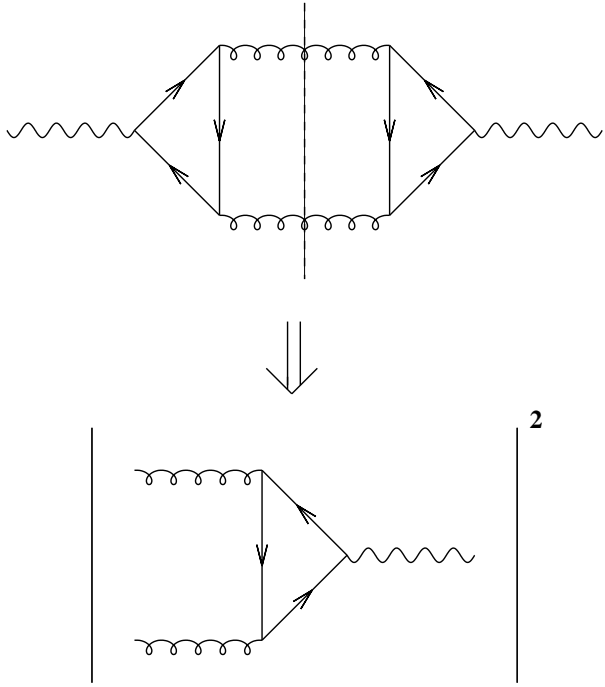}
    \caption{ The Full photon self-energy at three loop and the cut that is
evaluated in this paper. }
    \label{selfE}
  \end{center}
\end{figure}

We are now in a position to calculate the contribution made by the diagram of
Fig. \ref{2vert}
to the dilepton spectrum emanating from a quark gluon plasma. 
To achieve this aim we choose
to calculate the imaginary part of the photon self-energy as represented by the diagram of
Fig. \ref{selfE}. In the previous section we wrote down expressions for 
${\mathcal T}^{\mu \rho \nu}(k-p,k;p)$. To write down the expression for the full 
self-energy we also need expressions for ${\mathcal T}^{\zeta \rho \eta}(k,k-p;-p)$. 
A simple
analysis consisting of essentially reversing the direction of the internal momentum
$\vec{q}$ leads us to the result that ${\mathcal T}^{\zeta \rho \eta}(k,k-p;-p) = 
{\mathcal T}^{\eta \rho \zeta}(k-p,k;p)$. 

We may write down the full expression for the photon self-energy as

\begin{equation}
\Pi^{\rho}_{\rho} = \frac{1}{\beta} \sum_{k^{0}} \int \frac{d^{3}k}{(2\pi)^{3}} 
{\mathcal D}_{\eta \mu}(k) {\mathcal T}^{\mu \rho \nu} (k-p,k;p) 
{\mathcal D}_{\nu \zeta}(k-p){\mathcal T}^{\zeta \rho \eta}(k,k-p;-p). 
\end{equation}

\noindent
We perform this calculation in the Feynman gauge for the gluons, thus

\begin{equation}
{\mathcal D}_{\eta \mu}(k) = \frac{g_{\eta \mu}}{k^{2}}.
\end{equation}

\noindent 
We calculate in the limit of photon three momentum $\vec{p}=0$. We now shift 
$k \rightarrow
k+p$. Notice that, in the limit $\vec{p}=0$, this only implies shifting $k^{0}$ by $p^{0}$.
This is followed by switching $k^{0} \rightarrow -k^{0}$.
Note that neither of these operations has any effect on the full nature of the integral
as $k^{0}$ is summed over all even discrete frequencies. With this we now
obtain the full self-energy as 

\begin{eqnarray}
\Pi^{\rho}_{\rho} &=& e^{2} g^{4} \frac{32}{9} {\mathcal C}_{\alpha \beta \gamma}
^{\alpha^{\prime} \beta^{\prime} \gamma^{\prime} } 
\int \frac{ d^{3}k  d^{3}q dq^{\prime 3}  }
{(2\pi)^{9}} \frac{1}{\beta} \sum_{k^{0}} \sum_{s_{1}s_{2}s_{3} 
s_{1}^{\prime}s_{2}^{\prime}s_{3}^{\prime} } 
\frac{1}{(k^{0})^{2} - k^{2}}
\frac{1}{(p^{0}-k^{0})^{2} - k^{2}} \nonumber \\
& & \times \left[ 
 \frac{\widehat{(q-k)}_{s_{1}}^{\alpha}
 \hat{q}_{s_{2}}^{\beta} 
\hat{q}_{s_{3}}^{\gamma}
\{ s_{2} s_{1} \Delta \tilde{n}( E_{q},s_{3}\mu ) - 
s_{2} s_{3} \Delta \tilde{n}( E_{q-k},s_{1}\mu ) \} }{ 
( -k^{0}+p^{0} + s_{1}E_{q-k} - s_{3}E_{q} )
( -k^{0} + s_{1}E_{q-k} - s_{2}E_{q} ) } 
  \right] \nonumber \\
 & & \times \left[ 
 \frac{\widehat{(q^{\prime}-k)}_{s_{1}^{\prime},\alpha^{\prime}}
 \hat{q^{\prime}}_{s_{2}^{\prime},\beta^{\prime}} 
\hat{q^{\prime}}_{s_{3}^{\prime},\gamma^{\prime}}
\{ s_{2}^{\prime} s_{1}^{\prime} \Delta \tilde{n}( E_{q^{\prime}},s_{3}^{\prime}\mu ) - 
s_{2}^{\prime} s_{3}^{\prime} \Delta \tilde{n}( E_{q^{\prime}-k},s_{1}^{\prime}\mu ) \} }{ 
( -k^{0}+p^{0} + s_{1}^{\prime}E_{q^{\prime}-k} - s_{3}^{\prime}E_{q^{\prime}} )
( -k^{0} + s_{1}^{\prime}E_{q^{\prime}-k} - s_{2}^{\prime}E_{q^{\prime}} ) } 
  \right]. \label{fselfE}
\end{eqnarray}

Where the multiplicative factor ${\mathcal C}_{\alpha \beta \gamma}
^{\alpha^{\prime} \beta^{\prime} \gamma^{\prime} } $ is the result of the 
contraction of Lorentz indices appearing in the two traces of 6 $\gamma$ matrices 
and the intervening factors of the metric. In the above expression, we have also assumed the
presence of three massless flavours of quarks. For simplicity the chemical potential 
is assumed to be the same for all three flavours. 
In order to calculate the differential rate of back to back dileptons we need to evaluate
the imaginary part of this diagram. This may be obtained in two ways: the conventional
method \cite{kap89} involves converting the sum over discrete frequencies into a contour
integral, followed by evaluation of the contour integral by summing over its residues and
then looking for poles and branch cuts in the final expression in terms of $p^{0}$ by
analytically continuing $p^{0}$ onto the real axis. Ten years ago Braaten, Pisarski, and
Yuan presented an identity \cite{bra90} which elegantly achieved the end result of this 
procedure for a {\it fermionic} Matsubara frequency $k^{0}$ . In the following we 
will refer to this formula as the BPY formula. It is given, 
generally, in terms of products of arbitrary
functions of discrete imaginary frequencies as (for completeness, a brief proof of the BPY
formula for a {\it Bosonic} frequency $k^{0}$ in terms of the method of \cite{kap89} 
is presented in appendix D)

\begin{eqnarray}
\mbox{Disc}T \sum_{k^{0}} f_{1}(k^{0})f_{2}(p^{0}-k^{0}) &=& 2 \pi i (1-e^{E/T})
\int_{-\infty}^{+\infty}
d\omega \int_{-\infty}^{+\infty} d\omega^{\prime} \nonumber \\
& & n(\omega) n(\omega^{\prime}) 
\delta(E-\omega - \omega^{\prime}) \rho_{1}(\omega) \rho_{2}(\omega^{\prime}), \label{bpy} 
\end{eqnarray}
 
\noindent
where $\rho_{1}(\omega)$,$\rho_{2}(\omega^{\prime})$ are the spectral densities of 
$f_{1}(z)$,$f_{2}(z)$. Note that Eq. (\ref{fselfE}) is precisely in the form required
by Eq. (\ref{bpy}), with 

\begin{eqnarray}
f_{1}(k^{0}) &=&   \frac{1}{[k^{0}-(s_{1}x-s_{2}q)]
[k_{0}-(s_{1}^{\prime}x^{\prime}-s_{2}^{\prime}q^{\prime})]
[(k^{0})^{2} - k^{2}]} \nonumber \\
f_{2}(p^{0}-k^{0}) &=& \frac{1}{[p^{0}-k^{0}-(s_{1}x-s_{3}q)]
[k_{0}-(s_{1}^{\prime}x^{\prime}-s_{3}^{\prime}q^{\prime})]
[(p^{0}-k^{0})^{2} - k^{2}]},
\end{eqnarray}
  
\noindent where, for brevity we have  $x(x^{\prime})= E_{q-k}(E_{q^{\prime}-k})$ and 
$q(q^{\prime})=E_{q}(E_{q^{\prime}})$. From here, one may write down the spectral densities
of the two functions by mere inspection:

\begin{eqnarray}
\rho_{1}(\omega) &=&  \delta(\omega - k) \mbox{Res} 
[f_{1}(\omega = k)] +  \delta(\omega + k) \mbox{Res} 
[f_{1}(\omega = -k)] \nonumber \\
&+&
\delta(\omega - (s_{1}x-s_{2}q)) \mbox{Res}
[f_{1}(\omega = s_{1}x-s_{2}q)] \nonumber \\
&+& 
\delta(\omega - (s_{1}^{\prime}x^{\prime}-s_{2}^{\prime}q^{\prime}))
\mbox{Res}
[f_{1}(\omega = s_{1}^{\prime}x^{\prime}-s_{2}^{\prime}q^{\prime})] 
  \nonumber \\
\rho_{2}(\omega^{\prime}) &=& \delta(\omega^{\prime} - k) \mbox{Res} 
[f_{1}(\omega^{\prime} = k)] +  \delta(\omega^{\prime} + k) \mbox{Res} 
[f_{1}(\omega^{\prime} = -k)] \nonumber \\
&+&
\delta(\omega^{\prime} - (s_{1}x-s_{3}q)) \mbox{Res}
[f_{1}(\omega^{\prime} = s_{1}x-s_{3}q)]  \nonumber \\ 
&+&
\delta(\omega^{\prime} - 
(s_{1}^{\prime}x^{\prime}-s_{3}^{\prime}q^{\prime}))
\mbox{Res}
[f_{1}(\omega^{\prime} = s_{1}^{\prime}x^{\prime}-
s_{3}^{\prime}q^{\prime})]. \label{spden}
\end{eqnarray}

\noindent
In the above equation Res$[f(\omega)]$ stands for the residue of the function $f$ at 
$\omega$. In the language of BPY \cite{bra90}, the first two terms of the two
spectral densities are the pole terms. The next two are the cut terms as
they contain $q$ and $x$; variables which will get integrated over before
the $k$ integration. There are, of course, other multiplicative factors 
which
depend on $k,x,\mbox{ and }q$ which have not been expressly written 
down in
Eq. (\ref{spden}). 

We now take the product of the two spectral densities as given in
 Eq. (\ref{spden}). Each combination of delta functions gives us a
 different cut of the diagram. For instance combining any of the pole
 terms from the two spectral densities gives us the cut shown in Fig. \ref{selfE} 
 (and thus, the resulting
 cross section for the second diagram ). This represents the process of 
gluon-gluon to $\mbox{e}^{+}\mbox{e}^{-}$. This is a new process which to our
 knowledge has never been discussed before. The other possible cuts represent essentially
 finite density contributions to other known processes of dilepton production:
  two loop contributions to $q\bar{q}\rightarrow \gamma^{*}$, one loop
 contributions to $qg\rightarrow q\gamma^{*}$, and $qq\rightarrow qq\gamma^{*}$.

In the following we focus exclusively on the first process i.e., $gg
\rightarrow \gamma^{*}$. The reasons for this are twofold. The first is
the fact that this process is unique and resembles nothing at 
zero density; all the other processes are extra contributions to 
processes which are known and have been calculated up to two-loop level
at finite temperature and zero density \cite{aur98,aur99,kap00}.  
Secondly, all the other diagrams have at least one internal gluon line
which is weighted by Bose Einstein statistics, and hence, these
diagrams will display infrared divergence. To cure this divergence one 
can replace the bare gluon propagator with a resumed HTL propagator. These 
rather involved calculations deserve a
separate treatment. In contrast, all the internal lines of the first
process are fermions and hence show no infrared divergence.   

\vspace{1cm}

\section {THE CALCULATION}    

We now apply the BPY formula with only the pole-pole terms from
Eq. (\ref{spden}). The various delta function combinations that are
encountered can be schematically written as

\begin{eqnarray}
\delta(E-\omega-\omega^{\prime})\rho_{1}\rho_{2} &=& \delta(E-\omega-\omega^{\prime}) \Bigg[ 
\delta(\omega-k)\mbox{Res}[f_{1}(\omega=k)] 
\delta(\omega^{\prime}-k)\mbox{Res}[f_{2}(\omega^{\prime}=k)]  
\nonumber \\
&+& \delta(\omega-k)\mbox{Res}[f_{1}(\omega=k)] 
\delta(\omega^{\prime}+k)\mbox{Res}[f_{2}(\omega^{\prime}=-k)] 
\nonumber \\
&+& 
\delta(\omega+k)\mbox{Res}[f_{1}(\omega=-k)] 
\delta(\omega^{\prime}-k)\mbox{Res}[f_{2}(\omega^{\prime}=k)] 
\nonumber \\ 
&+&
\delta(\omega+k)\mbox{Res}[f_{1}(\omega=-k)] 
\delta(\omega^{\prime}+k)\mbox{Res}[f_{2}(\omega^{\prime}=-k)] \Bigg].
\end{eqnarray}

\noindent

Note that only the first term (i.e., gluon-gluon annihilation) survives. In
this term $\omega=\omega^{\prime}=k=E/2$. Thus, this contribution to the
discontinuity of the photon self-energy may be given as 

\begin{eqnarray}
\mbox{Disc}[\Pi^{\rho}_{\rho}] &=& e^{2} g^{4} \frac{64}{9}\pi i (1-e^{E/T})
{\mathcal C}_{\alpha \beta \gamma}
^{\alpha^{\prime} \beta^{\prime} \gamma^{\prime} } 
\int 
\frac{ 4\pi k^{2} 
 dq dq^{\prime} q^{2} q^{\prime 2} 
 d\theta d\theta^{\prime}
 \sin{\theta} \sin{\theta^{\prime}}  }
{(2\pi)^{7}}
\sum_{s_{1}s_{2}s_{3} 
s_{1}^{\prime}s_{2}^{\prime}s_{3}^{\prime} } 
\frac{1}{4k^{2}}
 \nonumber \\
& & \times \left[ 
 \frac{\widehat{(q-k)}_{s_{1}}^{\alpha}
 \hat{q}_{s_{2}}^{\beta} 
\hat{q}_{s_{3}}^{\gamma}
\{ s_{2} s_{1} \Delta \tilde{n}( E_{q},s_{3}\mu ) - 
s_{2} s_{3} \Delta \tilde{n}( E_{q-k},s_{1}\mu ) \} }{ 
( k + ( s_{1}E_{q-k} - s_{3}E_{q} ))
( k - ( s_{1}E_{q-k} - s_{2}E_{q} )) } 
  \right] \nonumber \\
 & & \times \left[ 
 \frac{\widehat{(q^{\prime}-k)}_{s_{1}^{\prime},\alpha^{\prime}}
 \hat{q^{\prime}}_{s_{2}^{\prime},\beta^{\prime}} 
\hat{q^{\prime}}_{s_{3}^{\prime},\gamma^{\prime}}
\{ s_{2}^{\prime} s_{1}^{\prime} \Delta \tilde{n}( E_{q^{\prime}},s_{3}^{\prime}\mu ) - 
s_{2}^{\prime} s_{3}^{\prime} \Delta \tilde{n}( E_{q^{\prime}-k},s_{1}^{\prime}\mu ) \} }{ 
( k + ( s_{1}^{\prime}E_{q^{\prime}-k} - s_{3}^{\prime}E_{q^{\prime}} ))
( k - (s_{1}^{\prime}E_{q^{\prime}-k} - s_{2}^{\prime}E_{q^{\prime}} )) } 
  \right]. \label{imselfE}
\end{eqnarray}

\noindent
As stated before, the multiplicative factor ${\mathcal C}_{\alpha \beta \gamma}
^{\alpha^{\prime} \beta^{\prime} \gamma^{\prime} } $ is the result of the 
contraction of Lorentz indices appearing in the two traces of 6 $\gamma$ matrices 
and the intervening factors of the metric. Essentially, this serves the purpose of the average over initial spins. 
We basically now have four terms to integrate (there is no mixed term between ($q,\theta$)
and ($q^{\prime},\theta^{\prime}$) ). Each term is classified by its fermionic 
distribution function. Now we may shift the variable 
$\vec{q} \rightarrow \vec{q} + \vec{k}$ in the second and fourth fermionic distribution
functions, and the $\theta$ and $\theta^{\prime}$ integrals become simple and can be done
analytically. The two $q$ and $q^{\prime}$ integrations are then done numerically.  

The differential production rate for pairs of massless leptons with total
energy $E$ and and total momentum $\vec{p}=0$ is given in terms of the discontinuity in the
photon self-energy as \cite{gal91} 

\begin{equation}
\frac{dW}{dEd^{3}p}(\vec{p}=0) = \frac{\alpha}{12\pi^{3}}\frac{1}{E^{2}}
\frac{1}{1-e^{E/T}}\frac{1}{2\pi i} \mbox{Disc}\Pi_{\rho}^{\rho}(0). \label{rate}
\end{equation}

\noindent 
Where $\alpha$ is the electromagnetic coupling constant. The rate of production of a hard
lepton pair with total momentum $\vec{p}=0$ at one-loop order in the photon self-energy 
(i.e., the Born term ) is given as 

\begin{equation}
\frac{dW}{dEd^{3}p}(\vec{p}=0) = \frac{\alpha^{2}}{6\pi^{4}}\tilde{n}(E/2 - \mu)
\tilde{n}(E/2 + \mu). \label{qqbar}
\end{equation}

\begin{figure}[htbp]
  \begin{center}
  \epsfxsize 108mm
  \epsfbox{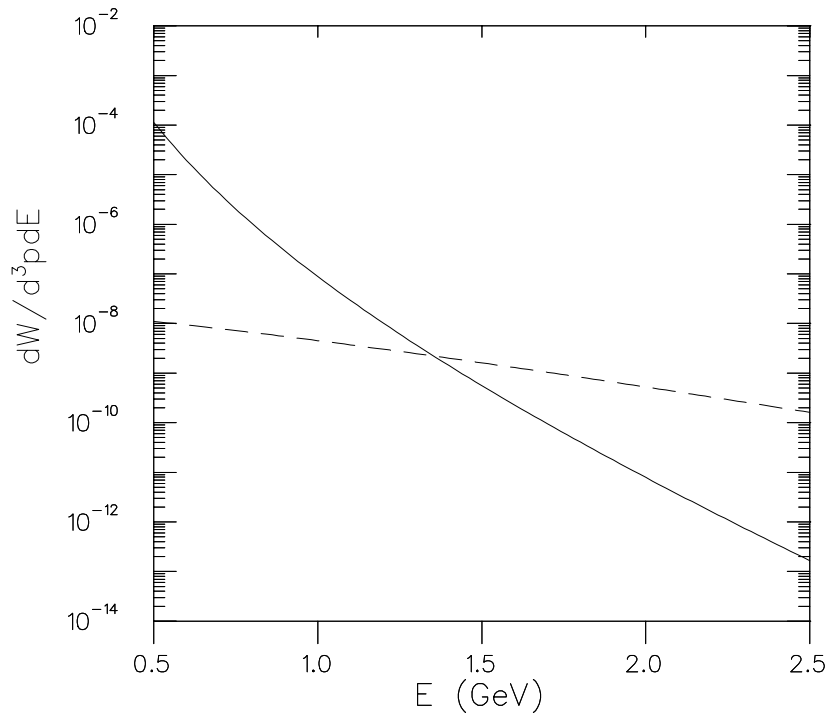}
\vspace{0.5cm}
    \caption{The differential production rate of back to back dileptons
from two processes. Invariant mass runs from 0.5 GeV to 2.5 GeV.
The dashed line represents the contribution from the process 
$q\bar{q}\rightarrow e^{+}e^{-}$. The solid line corresponds to 
the process $gg\rightarrow e^{+}e^{-}$. Temperature is 400 MeV. 
Quark chemical potential is 0.1T.  }
    \label{lowmu}
  \end{center}
\end{figure}

\begin{figure}[htbp]
  \begin{center}
  \epsfxsize 108mm
  \epsfbox{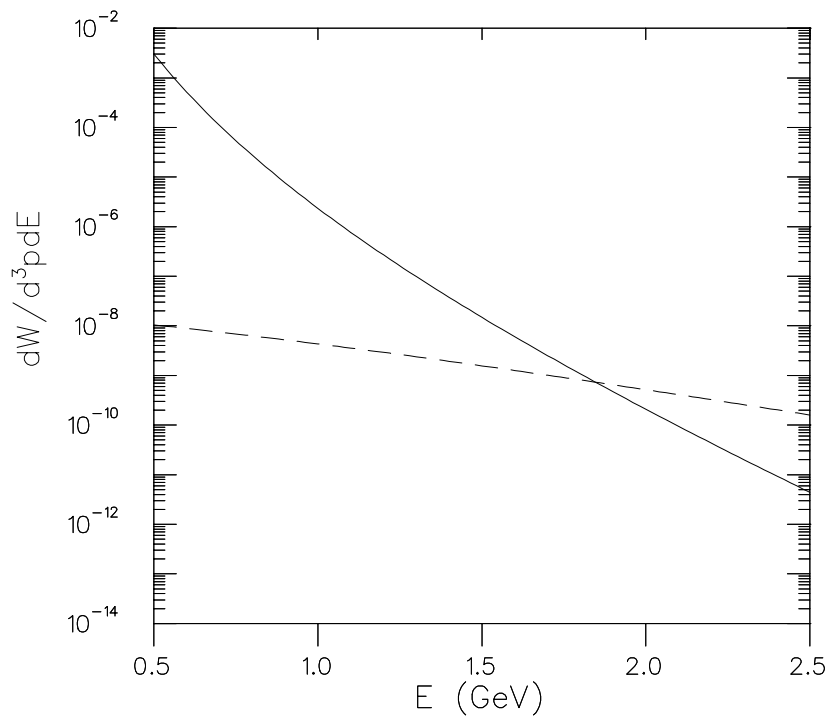}
    \caption{ Same as Fig. \ref{lowmu} but with $\mu=0.5T$ }
    \label{highmu}
  \end{center}
\end{figure}

 As mentioned before, initial temperatures of the plasma formed at RHIC and LHC 
have been predicted to lie in the range from 300-800 MeV \cite{wan96,rap00}.
 For this exploratory calculation 
we use a conservative estimate of $T=400\mbox{ MeV}$. To evaluate the effect
 of a finite chemical potential we perform the calculation  
with two extreme values of
chemical potential $\mu=0.1T$ (Fig. \ref{lowmu}) and $\mu=0.5T$(Fig. \ref{highmu})
\cite{gei93}.
The calculation, as stated before, is performed for three massless flavours of quarks. In this
case the strong coupling constant is (see \cite{kap00})
 
\begin{equation}
\alpha_{s}(T) = \frac{6\pi}{27ln(T/50MeV)}.
\end{equation}  

\noindent
 The differential rate for the production of dileptons with an invariant mass from 
 0.5 to $2.5\mbox{ GeV}$ is presented. On purpose, we avoid regions where the gluons become very soft.
  In the figures, the dashed line is the rate from tree level
 $q\bar{q}$ (Eq. (\ref{qqbar})); the solid line is that from the process 
 $gg\rightarrow e^{+}e^{-}$.  We note that in both cases the gluon-gluon process dominates
 at low energy and dies out at higher energy leaving the $q\bar{q}$ process dominant at
 higher energy. 

\vspace{1cm}

\section{DISCUSSIONS AND CONCLUSION} 

In this paper we have  performed a calculation to estimate the effects of a non-zero quark
chemical potential on the intermediate mass dilepton spectra. We have found that a new set
of diagrams may become important at finite temperature and finite density. These diagrams
lead to a new contribution to the photon self-energy at the 3-loop order. There are various
cuts to this diagram. Most of these result in finite density contributions, 
and/or higher order excess
contributions to well known processes. One of the cuts however represents an entirely new
process. The contribution of this diagram to the differential production rate of 
back-to-back (or $\vec{p}=0$) dileptons is estimated. 
This is then compared with the contribution
emanating from the tree level process of $q\bar{q}$ annihilation.
The rate from this new process is found to be larger than the simple tree level 
rate by orders of magnitude between a dilepton invariant mass of 
0.5 to 1.5 GeV. The reasons for this large magnitude are many. 
The gluon-gluon diagram is enhanced by 
the Bose-Einstein distribution function of the gluons. 
Also, at lower energy the $\Delta n$ factors
which represent the difference between the quark and antiquark distribution functions 
are comparable in size to the distribution functions themselves. 
There is, also, enhancement from the larger color factors of the gluons.
One also notes that as the energy (invariant mass) is increased, 
the difference between the
two distribution functions ($\Delta n$) decreases rapidly, i.e., as higher energies 
the distributions of quarks and antiquarks becomes less sensitive to the chemical 
potential. This is the main reason behind the sharp drop of the differential rate 
compared to the differential rate from quark-antiquark annihilation.   

These results demonstrate the importance of these finite $\mu$ processes on the
dilepton spectra emanating from a quark gluon plasma with a quark-antiquark asymmetry.
It is simple to note that this diagram is most sensitive to gluon number. The early
stages of the plasma have been predicted to be gluon dominated \cite{esk99}. 
The contribution from this
diagram should clearly shine in such an environment.    

The treatment in this rather exploratory work should, and will, be improved upon. Our goal
here was simply to establish the existence of a signal.
One may have desired, for example, 
that the calculation be extended to arbitrary $\vec{p}$. However,
similar extensions, even in the HTL approximation, are known to be rather involved
\cite{won92}. There are also the other diagrams which have yet to be computed. These
however suffer from the defect of having an internal boson line which will show 
an infrared
divergence. This bare propagator will have to replaced by a resumed HTL propagator, in 
the event that the momentum flowing through it becomes very small. These aspects, 
along with others, will be addressed elsewhere.

\vspace{1cm}

\section {ACKNOWLEDGMENT}

The authors wish to thank Y. Aghababaie, S. Das Gupta , F. Gelis,
 S. Jeon, D. Kharzeev,
 C. S. Lam and G. D. Mahlon for 
helpful discussions. A.M. acknowledges the generous support provided to 
him by McGill University through the Alexander McFee fellowship, the Hydro-Quebec
fellowship and the Neil Croll award.
This work was supported in part by the Natural
Sciences and Engineering Research Council of Canada and by { \it le fonds pour la
Formation de Chercheurs et l'aide \`{a} la Recherche du Qu\'{e}bec. }

\vspace{1cm}

\section {APPENDIX A}

\textbf{ Notation. }

Our notation is categorized by the explicit presence of an apparent Minkowski time 
$x^{0} = -i \tau$ and a momentum $q^{0} = i (2n+1) \pi T + \mu $. Our metric is 
$(1,-1,-1,-1)$.  For the case
of zero chemical potential our bosonic propagators have the same appearance as at zero
temperature, i.e., 

\begin{equation}
i\Delta(q) = \frac{i}{(q^{0})^{2} - |q|^2 }.
\end{equation}     

\noindent
The Feynman rules are also the same as at zero temperature, with the understanding that
we replace the zeroth component of the momentum by Eq. (\ref{q0}) for a fermion and
by an even frequency in the case of a boson. One may, in the case of zero chemical
potential, relate this to the familiar case of reference \cite{pis88} by noting that

\begin{equation}
\Delta(q) = \frac{1}{(q^{0})^{2} - |q|^2 } = \frac{-1}{(\omega_{n})^{2} + |q|^2 } 
= - \Delta_{E}(\omega_{n},q),
\end{equation}

\noindent
where $\Delta_{E}(\omega_{n},q)$ is the familiar Euclidean propagator presented in
the literature (\cite{pis88},\cite{kap89}). One may immediately surmise the form of
the non-covariant propagator $\Delta(|\vec{q}|,x^{0})$, the Fourier 
transform of which is the covariant propagator. 
 
\begin{eqnarray}
\Delta(q,q^{0}) &=& - \int_{0}^{\beta} d\tau 
e^{-i\omega_{n}\tau} \Delta_{E}(|\vec{q}|,\tau) \nonumber \\
&=& -i \int_{0}^{-i\beta} d x^{0} e^{iq^{0}x^{0}} \Delta_{E}(|\vec{q}|,\tau) 
\nonumber \\
&=& -i \int_{0}^{-i\beta} d x^{0} e^{iq^{0}x^{0}} \Delta(|\vec{q}|,x^{0}).
\end{eqnarray}

\noindent
In the presence of a finite chemical potential our full fermionic propagators become 

\begin{equation}
S(q,q^{0}) = ( \gamma^{\mu} q_{\mu} ) (-i) \int_{0}^{-i\beta} d x^{0} 
e^{ i( q^{0} - \mu )x^{0} }
\Delta_{\mu} (|\vec{q}|,x^{0}) ,
\end{equation}

\noindent where,

\begin{equation}
\Delta_{\mu} (|\vec{q}|,x^{0}) = \frac{1}{2E_{q}} \sum_{s} f_{s} ( E_{q} - s\mu )
e^{-isx^{0}( E_{q} - s\mu ) }.
\end{equation}

\noindent
In the above two equations $q^{0}= i(2n+1) \pi T + \mu$. 


\section {APPENDIX. B }

 \textbf{ Derivation of Eq. \ref{2g1pfv}. }

As stated in section III we start by substituting Eqs. (\ref{t1},\ref{t2},
\ref{t3}) in Eq. (\ref{2g1p}). Substituting in the first diagram we get, 

\begin{eqnarray}
T^{\mu \rho \nu} &=& \frac{1}{\beta} \sum_{n} \int \frac{d^{3}q}{2\pi^{3}}
 \frac{\delta_{ab}}{2} Tr[6\gamma] 
\sum_{s_{1},s_{2},s_{3}} 
\frac{ (q+p-k)_{s_{1} \alpha} q_{-s_{2} \beta} (q+p)_{s_{3} \gamma} }
{8E_{q+p-k}E_{q}E_{q+p}} 
\Bigg[ \int_{0}^{-i\beta} 
dx_{1}^{0} dx_{2}^{0} dx_{3}^{0}    \nonumber \\
& & e^{-i( q^{0} - \mu )x_{2}^{0}} 
    e^{i( q^{0} + p^{0} - k^{0} - \mu )x_{1}^{0}}
    e^{i( q^{0} + k^{0} - \mu )x_{3}^{0}}  \nonumber \\
& & \tilde{f}_{s_{1}} (E_{q+p-k},s_{1}\mu) 
    \tilde{f}_{s_{2}}^{\prime} ( E_{q},s_{2}\mu )  
     \tilde{f}_{s_{3}} ( E_{q+p},s_{3}\mu ) \nonumber \\
& & e^{-i s_{2}( E_{q} + s_{2}\mu )x_{2}^{0}}
    e^{-i s_{1}( E_{q+p-k} + s_{1}\mu )x_{1}^{0}} 
    e^{-i s_{3}( E_{q+p} + s_{3}\mu )x_{3}^{0}} \Bigg]. \nonumber \\ \label{appB} 
\end{eqnarray}

\noindent In our formalism $x^{0} = -i \tau$. Also, in Eq. (\ref{2g1p}) the zeroth 
components of the momenta are replaced by time derivatives e.g.,

\[
q_{0} = i\frac{d}{dx_{2}^{0}} + \mu, 
\]

\noindent
with the derivative acting on the $e^{-i(q^{0}-\mu)x_{2}^{0}}$. We then perform a by parts 
integration w.r.t. $x_{2}^{0}$, after which the derivative acts on the 
$e^{-i s_{2}( E_{q} + s_{2}\mu )x_{2}^{0}}$ term. After some algebraic manipulation and using 
the definition that $q_{s_{2} 0} = s_{2}E_{q}$, we obtain Eq. (\ref{appB}).  
We have the identity 

\begin{equation}
\frac{1}{\beta} \sum e^{iq^{0}(x_{1}^{0} - x_{2}^{0} + x_{3}^{0} )}
= \delta( \tau_{1} - \tau_{2} + \tau_{3} ) + 
(-1) \delta ( \tau_{1} - \tau_{2} + \tau_{3} - \beta ).
\end{equation}

\noindent Using the above identity we perform the Matsubara sum and 
then evaluate one of the 
 $\tau$ integrals using the two delta functions. Noting that 
$e^{p^{0} \beta} = 1$ we get 

\begin{eqnarray}
T^{\mu \rho \nu} &=&  \int \frac{d^{3}q}{2\pi^{3}} \frac{\delta_{ab}}{2} 
Tr[6\gamma]
\sum_{s_{1},s_{2},s_{3}}
\frac{ (q+p-k)_{s_{1} \alpha} q_{-s_{2} \beta} (q+p)_{s_{3} \gamma} }
{8E_{q+p-k}E_{q}E_{q+p}}
 \int_{0}^{-i\beta}  dx_{1}^{0} dx_{2}^{0} \nonumber \\
& & e^{i( p^{0} )x_{2}^{0}} 
    e^{-i( k^{0} )x_{1}^{0}} 
    \tilde{f}_{s_{1}} ( E_{q+p-k} , s_{1}\mu ) 
    \tilde{f}_{s_{2}}^{\prime} ( E_{q} , s_{2}\mu )  
    \tilde{f}_{s_{3}}( E_{q+p} , s_{3}\mu ) \nonumber \\
& &  e^{-ix_{2}^{0}( s_{2}E_{q} + s_{3}E_{q+p} )}
    e^{-ix_{1}^{0}( s_{1}E_{q+p-k} - s_{3}E_{q+p} )} 
\Bigg[ \Theta( \tau_{2} - \tau_{1} ) 
- \Theta( \tau_{1} - \tau_{2} ) e^{ \beta( -s_{3}E_{q+p} + \mu ) } \Bigg].
\end{eqnarray}

\noindent Noting that $e^{ \beta( -s_{3}E_{q+p} + \mu ) } = 
-\frac{ \tilde{f}_{-s_{3}}^{\prime}  ( E_{q+p} , s_{3}\mu )}
{ \tilde{f}_{s_{3}} ( E_{q+p} , s_{3}\mu ) }$ we may integrate 
$x_{2}^{0}$ making explicit use of the two Heaviside theta functions to obtain,

\begin{eqnarray}
T^{\mu \rho \nu} &=&  \int \frac{d^{3}q}{2\pi^{3}} \frac{\delta_{ab}}{2} 
Tr[6\gamma]
\sum_{s_{1},s_{2},s_{3}}
\frac{ (q+p-k)_{s_{1} \alpha} q_{-s_{2} \beta} (q+p)_{s_{3} \gamma} }
{8E_{q+p-k}E_{q}E_{q+p}}
 \int_{0}^{-i\beta}  dx_{1}^{0} \nonumber \\
& & e^{-ix_{1}^{0} ( k^{0} + s_{1}E_{q+p-k} -s_{3}E_{3} ) } 
    \tilde{f}_{s_{1}} ( E_{q+p-k} , s_{1}\mu ) 
    \tilde{f}_{s_{2}}^{\prime} ( E_{q} , s_{2}\mu ) \nonumber \\ 
& & \Bigg[  
 \tilde{f}_{s_{3}}( E_{q+p} , s_{3}\mu ) 
\frac{ e^{ i( -i\beta )( p^{0} - s_{2}E_{q} - s_{3}E_{q+p} )} 
- e^{ i( x_{1}^{0} )( p^{0} - s_{2}E_{q} - s_{3}E_{q+p} )} }
{ i( p^{0} - s_{2}E_{q} - s_{3}E_{q+p} ) } \nonumber \\
& & \mbox{} +  \tilde{f}_{-s_{3}}^{\prime} ( E_{q+p} , s_{3}\mu ) 
\frac{ e^{ i( x_{1}^{0} )( p^{0} - s_{2}E_{q} - s_{3}E_{q+p} )} - 1 }
{ i( p^{0} - s_{2}E_{q} - s_{3}E_{q+p} )} 
\Bigg]  \label{Tmng1}
\end{eqnarray}

\noindent We note that, 

\begin{eqnarray}
e^{ -\beta( s_{2}E_{q} + s_{3}E_{q+p} )} &=&
 e^{ -\beta s_{2}( E_{q} + s_{2}\mu  )}
 e^{ -\beta s_{3}( E_{q+p} - s_{3}\mu )} \nonumber \\
 &=& \frac{ \tilde{f}_{-s_{2}} ( E_{q} , s_{2}\mu ) }
 { \tilde{f}_{s_{2}}^{\prime}  ( E_{q} , s_{2}\mu ) }
\frac{ \tilde{f}_{-s_{3}}^{\prime}  ( E_{q+p} , s_{3}\mu ) }
 { \tilde{f}_{s_{3}} ( E_{q+p} , s_{3}\mu ) }.
\end{eqnarray}

\noindent Substituting the above equation in Eq. (\ref{Tmng1}), and noting
that $ \tilde{f}_{s} - \tilde{f}_{-s}^{\prime}  = s $, we may  
perform the $x_{1}^{0}$ integration to obtain

\begin{eqnarray}
T^{\mu \rho \nu} &=&  \int \frac{d^{3}q}{2\pi^{3}} \frac{\delta_{ab}}{2} 
Tr[6\gamma]
\sum_{s_{1},s_{2},s_{3}}
\frac{ (q+p-k)_{s_{1} \alpha} q_{-s_{2} \beta} (q+p)_{s_{3} \gamma} }
{8E_{q+p-k}E_{q}E_{q+p}}
\frac{1}{ p^{0} - s_{2}E_{q} - s_{3}E_{q+p} } \nonumber \\
& & \left[ \frac{ 
\tilde{f}_{s_{1}}( E_{q+p-k} , s_{1}\mu )
 \tilde{f}_{-s_{3}}^{\prime}  ( E_{q+p} , s_{3}\mu ) -   
\tilde{f}_{-s_{1}}^{\prime} ( E_{q+p-k} , s_{1}\mu ) 
\tilde{f}_{+s_{3}}( E_{q+p} , s_{3}\mu ) }
{ k^{0} + s_{1}E_{q+p-k} -s_{3}E_{q+p} }s_{2}  \right. \nonumber \\
& + & \left. s_{3} \frac{ 
\tilde{f}_{s_{1}}( E_{q+p-k} , s_{1}\mu )
 \tilde{f}_{s_{2}}( E_{q} , s_{2}\mu ) - 
\tilde{f}_{-s_{1}}^{\prime} ( E_{q+p-k} , s_{1}\mu ) 
\tilde{f}_{-s_{2}}^{\prime} ( E_{q} , s_{2}\mu )}
{ k^{0} - p^{0} + s_{1}E_{q+p-k} + s_{2}E_{q} }  \right].  
\end{eqnarray}

\noindent Expanding the $\tilde{f}$'s in terms of Fermi-Dirac distribution 
functions, we finally obtain the full expression for $T^{\mu \rho \nu}$  as 
(note: $s_{3}$ has been changed to $-s_{3}$ )

\begin{eqnarray}
T^{\mu \rho \nu} &=&  \int \frac{d^{3}q}{2\pi^{3}} \frac{\delta_{ab}}{2} 
Tr[6\gamma]
\sum_{s_{1},s_{2},s_{3}}
\frac{ (q+p-k)_{s_{1} \alpha} q_{-s_{2} \beta} (q+p)_{-s_{3} \gamma} }
{8E_{q+p-k}E_{q}E_{q+p}}
\frac{1}{ p^{0} - s_{2}E_{q} + s_{3}E_{q+p} } \nonumber \\
& & \left[ \frac{ (s_{1} + s_{3})/2 - 
s_{3}\tilde{n}( E_{q+p-k} -s_{1}\mu ) - 
s_{1}\tilde{n}( E_{q+p} -s_{3}\mu ) }
{ k^{0} + s_{1}E_{q+p-k} + s_{3}E_{q+p} } s_{2} \right. \nonumber \\
& - & \left. s_{3} \frac{ (s_{1} + s_{2})/2 - 
s_{2}\tilde{n}( E_{q+p-k} -s_{1}\mu ) - 
s_{1}\tilde{n}( E_{q} + s_{2}\mu ) }
{ k^{0} - p^{0}  + s_{1}E_{q+p-k} + s_{2}E_{q} } \right]. \label{2g1p1}
\end{eqnarray}

We, now, perform the same set of manipulations for the other Feynman diagram of
Fig. (\ref{2vert}) to obtain

\begin{eqnarray}
T^{\nu \rho \mu} &=&  \int \frac{d^{3}q}{2\pi^{3}} \frac{\delta_{ab}}{2} 
Tr[6\gamma]
\sum_{s_{1},s_{2},s_{3}}
\frac{ (q-p+k)_{s_{1} \alpha} q_{-s_{2} \beta} (q-p)_{-s_{3} \gamma} }
{8E_{q-p+k}E_{q}E_{q-p}}
\frac{1}{ p^{0} + s_{2}E_{q} - s_{3}E_{q-p} } \nonumber \\
& & \left[ \frac{ (s_{1} + s_{3})/2 - 
s_{3}\tilde{n}( E_{q-p+k} -s_{1}\mu ) - 
s_{1}\tilde{n}( E_{q-p} -s_{3}\mu ) }
{ k^{0} + s_{1}E_{q-p+k} + s_{3}E_{q-p} } \right. \nonumber \\
& - & \left. s_{3} \frac{ (s_{1} + s_{2})/2 - 
s_{2}\tilde{n}( E_{q-p+k} -s_{1}\mu ) - 
s_{1}\tilde{n}( E_{q-p} + s_{2}\mu ) }
{ k^{0} - p^{0}  + s_{1}E_{q-p+k} + s_{2}E_{q} } \right]. \label{2g1p2}
\end{eqnarray}

\noindent In the above two equations we have $Tr[6\gamma] = 
Tr[ \gamma^{\mu} \gamma^{\beta} \gamma^{\rho} 
\gamma^{\delta} \gamma^{\nu} \gamma^{\alpha}]$ in the first equation and  
$ = Tr[ \gamma^{\nu} \gamma^{\delta} \gamma^{\rho} 
\gamma^{\beta} \gamma^{\mu} \gamma^{\alpha}] $ in the second equation, note that both
these factors are the same. The only difference between the two equations is the sign
of $\vec{q}$. We now set $\vec{q} \rightarrow -\vec{q}$ in the three integration of
the second term. As a result $(q-p+k)_{s_{1} \alpha} \rightarrow 
-(q+p-k)_{-s_{1} \alpha} , q_{-s_{2} \beta} \rightarrow -q_{s_{2} \beta}$ and $
(q-p)_{-s_{3} \gamma} \rightarrow -(q+p)_{s_{3} \gamma}$. This is then followed by
setting $s_{1} \rightarrow -s_{1} , s_{2} \rightarrow -s_{2} , 
s_{3} \rightarrow -s_{3} $ in the summation over $s_{1},s_{2},s_{3}$.  With these
changes to Eq. (\ref{2g1p2}), we may now add Eqs. (\ref{2g1p1}) and 
(\ref{2g1p2}), and finally change $(s_{2}$, $s_{3}) \rightarrow (-s_{2}$, $-s_{3})$ 
to give Eq. (\ref{2g1pfv}).  
  
\vspace{1cm}

\section {APPENDIX. C }

\textbf{ Interdependence of the ${\mathcal T}^{\mu \rho \nu}$ integrals }

In this appendix we will analyse the (potentially 64) terms generated by 

\begin{equation}
{\mathcal J}^{\alpha \beta \gamma} = \frac{{\mathcal T}^{\mu \rho \nu}}
{Tr[\gamma^{\mu} \gamma^{\beta} \gamma^{\rho} 
\gamma^{\delta} \gamma^{\nu} \gamma^{\alpha}]}.
\end{equation}

\noindent
We note that the Lorentz indices affect, specifically, only part of the integrand
of ${\mathcal J}^{\alpha \beta \gamma}$, i.e., we may write 

\begin{equation}
{\mathcal J}^{\alpha \beta \gamma} = \int d\theta d\phi \sin{\theta} \Bigg[ 
I^{\alpha \beta \gamma}(\theta,\phi) \Bigg( \int dq q^{2} J(q,\theta,\mu) \Bigg)
\Bigg]. 
\end{equation}

\noindent
As noted in section III, performing the $\phi$ integration sets 44 of the 64 terms
to zero, the only surviving terms are those that carry $(\alpha, \beta, \gamma)$ in
some permutation of the combinations (0,0,0),(3,3,3),(0,0,3),(0,3,3), 
($\phi$ integration gives $2\pi$), and those of (0,1,1),(0,2,2),(3,1,1),(3,2,2), 
($\phi$ integration gives $\pi$). The $\mu$ dependence in $J$ only distinguishes between
terms with $\mu = 0$ and $\mu \not= 0$. On performing the $\phi$ integration we get 

\begin{equation}
{\mathcal J}^{\alpha \beta \gamma} = \int d\theta \sin{\theta} \Bigg[ 
I^{\alpha \beta \gamma}(\theta) \Bigg( \int dq q^{2} J(q,\theta,\mu) \Bigg)
F( \alpha, \beta, \gamma) \Bigg], 
\end{equation}
 
\noindent
where $J(q,\theta,\mu)= J_{0}\delta_{\mu,0} + J_{1}(1-\delta_{\mu,0})$, and  

\begin{eqnarray} 
F( \alpha, \beta, \gamma) &=& \pi \Big[ ( g_{\alpha,1}g_{\beta,1} + 
g_{\alpha,2}g_{\beta,2} )( g_{\gamma,0} - g_{\gamma,3} ) +
( g_{\alpha,1}g_{\gamma,1} + g_{\alpha,2}g_{\gamma,2} )(
g_{\beta,0} - g_{\beta,3} ) \nonumber \\
&+& 
( g_{\beta,1}g_{\gamma,1} + g_{\beta,2}g_{\gamma,2} )(
g_{\alpha,0} - g_{\alpha,3} ) \Big]
 + 2\pi \Big[ ( g_{\alpha,0}g_{\beta,0} + 
g_{\alpha,3}g_{\beta,3} )( g_{\gamma,0} - g_{\gamma,3} ) \nonumber \\
&-& 
g_{\alpha,0}g_{\gamma,0}g_{\beta,3} + g_{\alpha,3}g_{\gamma,3}g_{\beta,0} -
g_{\beta,0}g_{\gamma,0}g_{\alpha,3} + g_{\beta,3}g_{\gamma,3}g_{\alpha,0}\Big].  \nonumber 
\end{eqnarray}

We now look at the integrands of the $\theta$ integration: note for example that 

\[
I^{0 0 0} = s_{1}s_{2}s_{3} ;\mbox{ } I^{0 0 3} = s_{1}s_{2}\cos{\theta} ;\mbox{ } 
I^{0 3 0} = s_{1}\cos{\theta}s_{3}.
\]

\noindent
The reader may easily verify that this implies that

\begin{equation}
 {\mathcal J}^{003} = \frac{s_{3}}{s_{2}}{\mathcal J}^{030}.
\end{equation}

\noindent 
Also note that
\[
I^{033} = s_{1}\cos^{2}{\theta}  ; I^{011} = I^{022} = s_{1}\sin^{2}{\theta},
\]

\noindent 
which implies that

\begin{eqnarray}
 {\mathcal J}^{011} &=& {\mathcal J}^{022} \nonumber \\
 {\mathcal J}^{033} &+&  2{\mathcal J}^{011}= \frac{{\mathcal J}^{000}}{s_{2}s_{3}}.
\end{eqnarray}

\noindent 
Following the above method we may also prove that 

\begin{eqnarray}
 {\mathcal J}^{303} &=& \frac{s_{2}}{s_{3}}{\mathcal J}^{330} \nonumber \\
 {\mathcal J}^{311} &=& {\mathcal J}^{322} \nonumber \\
 {\mathcal J}^{131} &=& {\mathcal J}^{232} = {\mathcal J}^{113} = {\mathcal J}^{223}
 \nonumber \\
 {\mathcal J}^{333} &+& 2{\mathcal J}^{311} = \frac{{\mathcal J}^{300}}{s_{2}s_{3}}.
\end{eqnarray}

\noindent
The above 12 conditions reduce the number of independent 
${\mathcal J}^{\alpha \beta \gamma}$'s to 8. Thus, one only has to evaluate these 8 terms
and the others may be evaluated by the above mentioned conditions.
 
\vspace{1cm}

\section {APPENDIX. D }

\textbf{ Derivation of the BPY formula. }

In this section we present a discussion on the BPY formula. The basic aim is to evaluate the
quantity

\[
\mbox{Disc} \Bigg[ S = T \sum_{k^{0}} f_{1}(k^{0})f_{2}(p^{0}-k^{0}) \Bigg],  
\]

\noindent
where $p^{0}$,$k^{0}$  are both discrete frequencies. As stated in \cite{kap89}
the sum over the Matsubara frequencies can be converted into two
 contour integrals
over a complex $k^{0}$ ( Eq. (3.39) of \cite{kap89} ), i.e.,

\begin{eqnarray}
S &=& \frac{1}{2 \pi i}\int_{ i\infty - \epsilon }^{ -i\infty - \epsilon } dk^{0}
f_{1}(k^{0})f_{2}(p^{0}-k^{0})( \frac{1}{2} +
\frac{1}{e^{ \beta k^{0} } - 1 } )  \nonumber \\
&+&  \int_{ -i\infty + \epsilon }^{ +i\infty + \epsilon } dk^{0}
f_{1}(k^{0})f_{2}(p^{0}-k^{0})( \frac{1}{2} +
\frac{1}{e^{ \beta k^{0} } - 1 } ). \label{cont}
\end{eqnarray}

Now $f_{1}(z)$ and $f_{2}(z)$ both have discontinuities on the real axis, which
consists mostly of residues and branch cuts. Let the sum total of all such
discontinuities be set equal to a spectral density $\rho (\omega)$, i.e., 

\begin{equation}
2\pi i \rho (\omega) = 
\sum_{i} \delta ( \omega - p_{i} ) 2\pi i \mbox{Res}[f (z= \omega)] + 
\sum_{j} \Theta(\omega-l_{j})\Theta(u_{j}-\omega)\mbox{Disc}[f(z=\omega)], 
\end{equation}

\noindent
where $p_{i}$ represent the location of the poles of $f (z)$ on the real
axis ,and $u_{j}$ and $l_{j}$  represent the upper and lower bounds of the
branch cuts on the real axis. The presence of $p^{0}$ in the arguments of only
one of the functions in Eq. (\ref{cont}) separates the discontinuities of
the two functions. We have tacitly assumed that both these functions contain
integrable discontinuities. Thus the contour integration can be reduced to a sum
over residues and discontinuities of the two functions separately. 
Let us look at the first integral. The function $f_{1}(k^{0})$ 
($k^{0}$ now 
complex ) has discontinuities at $k^{0}=\omega (\mbox{ real})$. The
spectral density $\rho_{1}(\omega)$ is defined as 

\[
2\pi i \rho_{1}(\omega) = [f_{1}(z=\omega + i\epsilon) - 
f_{1}(z=\omega - i\epsilon)],
\]

\noindent
thus the total contribution of this term to the clockwise integral is 

\[
-\int_{-\infty}^{0} d\omega \rho_{1}(\omega) f_{2}(p^{0}-\omega)
\Bigg[ \frac{1}{2} + \frac{1}{ e^{\beta \omega} - 1 } \Bigg].
\]

\noindent
The function $g(k^{0})=f_{2}(p^{0}-k^{0})$ has discontinuities at 
$k^{0}= p^{0} - \omega (\mbox{ real})$. The spectral density now, is 

\[
 = \Big[ g(z=p^{0} - \omega + i\epsilon) 
- g(z= p^{0} - \omega - i\epsilon) \Big]
\]
 
\[
 = \Big[ f_{2}(z = \omega - i\epsilon) 
- f_{2}(z = \omega + i\epsilon) \Big] = - 2 \pi i \rho_{2}(\omega).
\]

\noindent
Thus the contribution from this term to the first clockwise integral  is

\[
\int_{0}^{\infty} d\omega^{\prime} \rho_{2}(\omega^{\prime}) f_{1}(p^{0}-\omega^{\prime})
\Bigg[ \frac{1}{2} + \frac{1}{ e^{\beta(p^{0}- \omega^{\prime})} - 1 } \Bigg].
\]

\noindent
Performing the above mentioned operation for the other contour integral we get,
  
\begin{eqnarray}
S &=& - \int_{-\infty}^{0} d\omega  \rho_{1}(\omega) f_{2}(p^{0}-\omega) 
\Bigg[ \frac{1}{2} + \frac{1}{ e^{\beta \omega} - 1 } \Bigg] 
+ \int_{0}^{\infty} d\omega^{\prime} f_{1}(p^{0}-\omega^{\prime}) 
\rho_{2}(\omega^{\prime}) 
\Bigg[ \frac{1}{2} + \frac{1}{ e^{\beta (p^{0} - \omega^{\prime} ) } - 1 } 
\Bigg] \nonumber \\
& &\!\!\!\!\!\!\!\!\!\!\!\!\!\!\!\!\!\!\!\!\!\!\!\!\!
\mbox{} - \int^{\infty}_{0} d\omega  \rho_{1}(\omega) f_{2}(p^{0}-\omega) 
\Bigg[ \frac{1}{2} + \frac{1}{ e^{\beta \omega} - 1 } \Bigg] 
+ \int^{0}_{-\infty} d\omega^{\prime} f_{1}(p^{0}-\omega^{\prime})  
\rho_{2}(\omega^{\prime}) 
\Bigg[ \frac{1}{2} + \frac{1}{ e^{\beta ( p^{0} - \omega^{\prime} )} - 1 } 
\Bigg]. \label{res}
\end{eqnarray}

\noindent
The two lines in Eq. (\ref{res}) display the contributions from the two
separate contour integrals of Eq. (\ref{cont}). These may be combined
together to give,

\begin{eqnarray}
S(p^{0}) &=& - \int_{-\infty}^{\infty} d\omega  
\rho_{1}(\omega) f_{2}(p^{0}-\omega) 
\Bigg[ \frac{1}{2} + \frac{1}{ e^{\beta \omega} - 1 } \Bigg]  \nonumber \\
&+& \int_{-\infty}^{\infty} d\omega^{\prime} f_{1}(p^{0}-\omega^{\prime}) 
\rho_{2}(\omega^{\prime}) 
\Bigg[ \frac{1}{2} + \frac{1}{ e^{\beta (p^{0} - \omega^{\prime} ) } - 1 } 
\Bigg].
\end{eqnarray}

Now, we start the operation of analytically continuing $p^{0}$ onto the real
axis. Note that the Bose factor of the second term will have a discontinuity when $p^{0}$ is
analytically continued onto the real axis. Thus the correct analytic continuation is
achieved by dropping the $p^{0}$ from the expression (as $\exp{\beta p^{0} }= 1 
\mbox{ when } p^{0} = -i2n \pi T $). 
We observe that in the first term of Eq. (\ref{res}) the only factor
 that may have a discontinuity is $f_{2}(p^{0}-\omega)$ and this only happens
 when $p^{0}-\omega = \omega^{\prime}$(a real number). This is only possible when
 $p^{0} \rightarrow E \pm i \epsilon$. Taking the discontinuity across the real
 axis in both terms of Eq. (\ref{res}), and combining them  together we get,
 
\begin{equation}
\mbox{Disc}[S(E)] =  -2\pi i \Bigg[ \int_{-\infty}^{\infty} d \omega 
 \int_{-\infty}^{\infty} d \omega^{\prime} \rho_{1}(\omega)
  \rho_{2}(\omega^{\prime}) \delta( E - \omega - \omega^{\prime} ) 
  \Bigg\{ \frac{1}{ e^{\beta \omega} - 1 } - \frac{1}{ e^{-\beta \omega^{\prime}} - 1 }
  \Bigg\} \Bigg].    
\end{equation}

\noindent We observe that the term in the curly brackets may be simplified:
\[
 \frac{1}{ e^{\beta \omega} - 1 } - \frac{1}{ e^{-\beta \omega^{\prime}} - 1 } =
 \frac{e^{\beta E} - 1}{ (e^{\beta \omega} - 1)(e^{-\beta \omega^{\prime}} - 1) },
\]

\noindent
thus we get back the BPY formula, i.e.,

\begin{eqnarray}
\mbox{Disc}T \sum_{k^{0}} f_{1}(k^{0})f_{2}(p^{0}-k^{0}) = 2 \pi i (1-e^{E/T})
\int_{-\infty}^{+\infty}
d\omega \int_{-\infty}^{+\infty} d\omega^{\prime} \nonumber \\
n(\omega) n(\omega^{\prime}) 
\delta(E-\omega - \omega^{\prime}) \rho_{1}(\omega) \rho_{2}(\omega^{\prime}).  
\end{eqnarray}




%

\end{document}